\documentstyle[epsf]{lamuphys}
\makeatletter
\let\chapter\hid@chapter
\makeatother
\begin{document}
\pagenumbering{arabic}
\title{Polarimetric Evidence of Non-Spherical Winds}

\author{Ant\^onio M\'ario\,Magalh\~aes\inst{1} and
Cl\'audia V.\,Rodrigues\inst{2}}

\institute{IAG, Univ. S\~ao Paulo, Caixa Postal 3386 - S\~ao Paulo
SP 01060-970 - Brazil \and INPE, Av. dos Astronautas, 1758 - S\~ao
Jos\'e dos Campos SP 12227-900 - Brazil }

\maketitle

\begin{abstract}
Polarization observations yield otherwise unobtainable information
about the geometrical structure of unresolved objects. In this
talk we review the evidences for non-spherically symmetric
structures around Luminous Hot Stars from polarimetry and what we
can learn with this technique. Polarimetry has added a new
dimension to the study of the envelopes of Luminous Blue
Variables, Wolf-Rayet stars and B[e] stars, all of which are
discussed in some detail.\footnote{Invited review to appear in IAU
Coll. 169, {\it Variable and Non-spherical Stellar Winds in Luminous Hot
Stars}, eds. B. Wolf, A.
Fullerton and O. Stahl (1999).}
\end{abstract}
\section{Introduction}
In the past few years there has been mounting evidence that the mass loss in
Luminous Hot Stars (LHS) is non-spherically symmetric and this meeting is in
fact a
testimony to that. In addition, the abundance of free electrons in the winds of
such objects makes Thomson scattering an important opacity source. This
combination of asymmetry and scattered (hence polarized) light may result in an
observed degree of polarization in the radiation we detect from LHS.
Polarization observations carry then great potential to explore the environment
of LHS.

In this talk we review the evidences for non-spherically symmetric
structures around LHS from polarimetry and what we can learn from
such data about the physics of such structures. Recent related
reviews include those of Bjorkman (1994) and Schulte-Ladbeck
(1997). Several talks in this conference also have direct bearing
on the topic (K. Bjorkman, Brown and Ignace, Eversberg et al.,
Rodrigues and Magalh\~aes, Schulte-Ladbeck et al.).

\section{Some Polarimetry Basics}

One great asset of polarization observations is that they yield
diagnostics related to the geometrical structure of unresolvable
objects. Generally, it can be said that the polarization is the
ratio between the scattered flux and the total flux from the
object. The polarization from a stellar envelope will depend in
detail on the density and geometrical distribution of matter
around the star (e.g., \cite{W96}). Techniques for measuring
polarization in the UV-optical-IR have greatly advanced in recent
years (\cite{rw96}; \cite{mag96}).

The polarimetric wavelength dependence may be modified by any
competing opacity and any unpolarized, diluting light from the
star and/or wind. Examples include hydrogen bound-free and
free-free opacities as well as line opacity such as from iron.
Hydrogen recombination line emission tends to decrease the
polarization across corresponding features, such as Balmer lines.
All this provides valuable wind diagnostics. Dust scattering can
also play a role in the outskirts of evolved objects. The
wavelength dependence of single dust scattering depends on the
nature of the grains and their size.

While this review will be concerned mostly with linear polarization, circular
polarization may also in principle arise from processes such as multiple dust
scattering in an envelope or magneto-emission from stellar spots. Electron
scattering produces no circular polarization by itself.

Intrinsic polarization may be detected from the time variability
of the observed polarization. In addition, the scatter of the data
points in the Q-U diagram (Q = P.cos${(2\theta)}$ and U =
P.sin${(2\theta)}$, where P=percent polarization,
$\theta$=position angle) will tell whether there is a preferred
plane of symmetry or not. Binary stars where the scattering
envelope surrounds one of them will show up as loops in the Q-U
diagram (\cite{br78}). Intrinsic polarization may also show up
through spectropolarimetry. If the observed polarization varies
across a line, such as H$\alpha$, the vector difference in the Q-U
plane of the continuum and line polarizations will provide the
position angle (PA) of the intrinsic polarization (e.g.,
\cite{sl92}).

\section{Observations of Luminous Hot Stars}

\subsection{Luminous Blue Variables}

Luminous Blue Variables (LBV) represent an intermediate stage between OB and WR
stars (\cite{M96}). Direct evidence for asymmetric outflows comes from imaging
(cf. Nota, these proceedings). In this case, spectropolarimetry has been used to
probe mass loss on small spatial scales.


The {\bf P Cyg} nebula has been resolved by direct imaging by
\cite{lz87}. P Cyg shows stochastic changes in its optical linear
polarization (\cite{h85}), with night to night changes of 0.2\%
and 6$^o$ in the polarization degree and PA, respectively.

\cite{t91a} have obtained spectropolarimetry of P Cyg for 20
nights during the 1989-1990 season. The observed polarization
showed no preferred plane, consistent with random ejections of
matter from the star. No correlation between increased line
emission and polarization was observed. This was interpreted as a
result from the time lag between these events, since about
40$^{d}$ are required for a mass ejection to travel out to a
distance of about 3 R$_{*}$, within which the polarization is
thought to be produced.

Further constraints on P Cygni's envelope came from UV spectropolarimetry
with WUPPE (\cite{t91b}). A broad dip in the polarization around
2600-3000\AA\ suggested the existence of an absorptive opacity by FeII
lines in the envelope. High resolution imaging of P Cyg (Nota et al. 1995; Nota,
these proceedings) shows that the structure of the envelope is indeed
clumpy, nicely consistent with the structures seen much closer to the
star in the polarimetry data.

\label{r127}

The Large Magellanic Cloud LBV {\bf R 127} has been observed for
spectropolarimetry by Schulte-Ladbeck et al. (1993). The intrinsic
polarization, indicated by the line effect at H$\alpha$, showed a
level around 1.5\% and was suggestive of electron scattering with
possible FeII depression from within the envelope. The
polarization was variable but with PA values restricted within a
'cone', with the interstellar value as apex, in the Q-U diagram.

The observed nebula (\cite{c93}) is about 2\,pc in size and
$\approx$ 10$^{4}$ yr old. There are symmetric enhancements in the
(coronographic) image along a direction $\approx$ 90$\degr$ from
the polarization PA value. The suggested geometry for R127
(\cite{sl93}; Fig. 1) is then that of a mass ejection in a
preferred plane. The present geometry (from imaging) is defined by
events taking place very close to the star (from polarimetry).

\begin{figure}[t]
\begin{center}
\epsfxsize=4cm
\leavevmode\epsffile{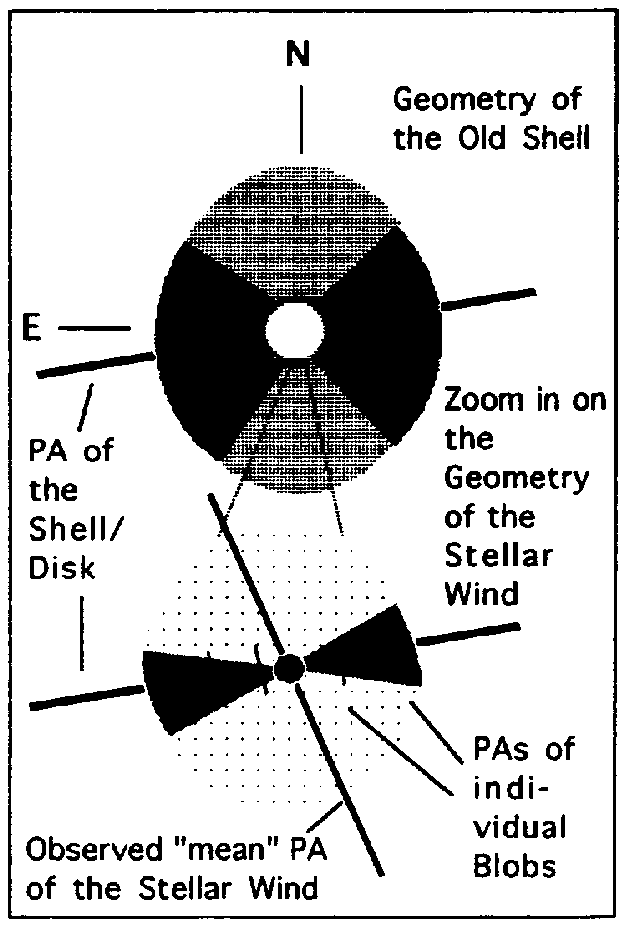}
\caption[]{Geometry of circumstellar wind of R127
(\cite{sl93}).}
\end{center}
\end{figure}


A few other LBV have been observed polarimetrically. In AG Car (\cite{l94};
\cite{sl94b}),
the geometry of the nebula shows an alignment with the PA derived from
spectropolarimetry, with broad, polarized
wings across H$\alpha$ suggesting electron
scattering. In HR Car (\cite{c95}) the PA from imaging and
that from polarimetry are actually the same, about 30$^o$. However, we note
that, according to
\cite{w97}, the bipolar nebula has actually its axis
at PA$\approx$125\,$\degr$ and the imaging and polarimetry
data are again consistent. Further monitoring of these objects
to confirm the ejections in a preferred plane would be highly desirable.

In summary, polarimetry indicates that LBV may show either
stochastic ejections (P Cyg) or, more commonly, a preferred
plane for mass loss. In any case, the geometry present in the observed
nebulae is already present in (and presumably imposed by) the wind very
close to the star. Possible sources for this density contrast have
been conjectured by \cite{n95} but it is not possible yet to discern
among them.

\subsection{Wolf-Rayet Stars}

Wolf-Rayet (WR) stars are the polarimetrically best studied class
among the LHS (e.g., \cite{r89}; \cite{mr91}; \cite{d92}). For
(presumed) {\bf single} WR stars, there is a range in the observed
variations of optical linear polarization: (a) WN stars vary more
than WC ones in a given subclass; (b) Cooler sub-types (i.e.,
slower winds) vary more, although a few ($\approx$20\%) WR show no
variability; (c) Polarization variations have time scale of days
and are wavelength independent; (d) Most WR show no preferred
plane, but there are a few exceptions. Intraday variability is
still poorly known.

For {\bf binary} WR stars, cyclic variations of polarization with
binary phase are often seen. This is due to the O-star light
scattered off the dense WR wind. Mass loss rates can be derived
(\cite{stl88}) as well as the inclination of the systems
(\cite{br78}), providing important information about WR masses.

Circular polarimetry has been looked for in EZ Cma (\cite{r92}) with
negative results, suggesting that the star does not show activity related to
strong magnetic fields.

\cite{h98} performed a spectropolarimetric survey of 16 WR. Their data are
consistent with
a distribution of intrinsic polarizations biased towards small values,
with only $\approx$ 20\% of stars with P $\ge$ 0.3\%. Radiative transfer
models suggest equator-to-pole density contrast of 2-3.
Combining their results with literature data, for a total of 29 stars,
the 5 known objects with 'line effects' cluster around the
high mass loss \&
luminosity part of the $\dot{M}$-L diagram (Fig. 2). Also, the $\dot{M}$
values from radio and optical are in good agreement, suggesting that
the wind structures have density contrast independent of radius.

\begin{figure}
\begin{center}
\leavevmode
\epsfxsize=5.5cm
\epsfbox{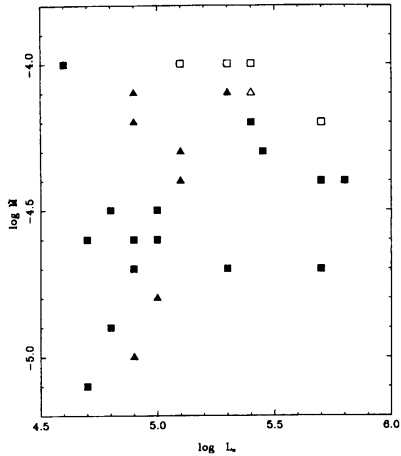}
\end{center}
\caption[]{Spherical (filled symbols) and
non-spherical (open symbols) WR stars in the $\dot{M}$-L diagram
(Harries et al. 1998).}
\end{figure}

The results of \cite{h98} seem to suggest that the global wind
asymmetries in WR winds arise only in the fastest rotators (\cite{i96}).
Specially in view of the distribution of 'line effect' stars in the $\dot{M}$-L
diagram, we feel that this is also supported by the fact that rotating stars
evolve towards higher luminosity (\cite{f96}).

As the O component screens the WR envelope in an eclipse, the
observed polarization may change dramatically and it can be used to
model the WR wind (e.g., \cite{stl93}; \cite{rm95}). Spectropolarimetry across
eclipses holds also great potential for probing the ionization structure of the
wind.

\subsection{B[e] Supergiants}

These objects show evidence of a two-component wind: a hot, fast
polar wind and a denser, slow equatorial wind (see Zickgraf, these
proceedings; \cite{ar94}). Magalh\~aes (1992) showed that the
Magellanic B[e] supergiants do present intrinsic polarization,
lending further support to the model put forward by Zickgraf et
al. In addition, the higher intrinsic polarization values were all
associated with objects spectroscopically found to be edge-on. The
polarization of these systems, $P_{edge-on}$, correlated some with
the average electron density $N_{e}$ of the envelopes but it
correlated somewhat better with the IR [K-L] dust excesses.

\cite{m92} pointed out however that in the IR we tend to detect
the larger grains, which are poor scatterers in the optical and
might not polarize; instead, electrons closer to the star might be
operative.  Interestingly, the $P_{edge-on}-N_{e}$ correlation of
Magalh\~aes (1992) with the AV16/R4 (a binary, \cite{z96b}) point
removed becomes actually the tightest one. Spectropolarimetry of
the most highly polarized object (S22, \cite{slc93}) showed that
electron scattering is indeed present, at least for that object.
Monte Carlo scattering models (\cite{mel99}) suggest that {\it
homogeneous} disks fit the polarization data for B[e] stars well.
This is consistent with the very slow winds observed and modeled
by Zickgraf et al. (1996a) from spectroscopic data, providing
another interesting link between the different types of
observations.

Three Magellanic B[e] stars have shown variability
in polarization (S22, \cite{slc93}; S18, \cite{sl94a}) and photometry and
spectroscopy (R4, \cite{z96b})
similarly to LBV stars. While further scrutiny may show others to be variable
too, \cite{g95} have shown that the B[e]
class actually extends to luminosities much lower (log L/L$_{\sun}\approx
4$) than their supergiant counterparts (log L/L$_{\sun}\approx 5.5-6.0$).

\subsection{Other Objects}

\cite{ln87} showed that OB stars have intrinsic polarization. An
on-going spectropolarimetric survey of OB supergiants is being
conducted by Karen Bjorkman (\cite{b94}). The observed random PA
values suggest that instabilities in an otherwise spherical wind
(rather than in a disk) are the cause of the variations. The less
luminous Be stars, which show disks, are discussed by K. Bjorkman
elsewhere in these proceedings.

Another class of LHS is the Ofpe/WN9 stars, of which ten or so are known in the
Magellanic Clouds. They may be O stars in transition to the WR
Stage that experience an LBV stage with Ofpe/WN9 characteristics in quiescence
(\cite{cs97}). R127 (section \ref{r127}) has
actually become an LBV from an Ofpe/WN9 object
(\cite{s83}). \cite{p97} showed that HDE 269445
has a two component wind. Undoubtedly this class as a whole would be a prime
target for polarimetric studies.

\section{Conclusions}

Imaging and spectropolarimetry data indicate that non-spherically
symmetric winds about LHS are the norm. In addition to the
suggested systematic observations, other new polarimetric
techniques, such as using the Hanle effect in the UV for sensitive
detection of magnetic fields (\cite{n96}) look promising. In
addition, the new generation of large aperture telescopes such as
Gemini and VLT will offer polarimetric capabilities that will be
important particularly for the study of objects in the Magellanic
Clouds. At the same time, detailed envelope modeling is just
becoming possible especially due to Monte Carlo techniques,
providing an important feedback on theoretical models. The next
few years are bound to witness the coming of age of polarimetry of
Luminous Hot Stars and the tapping of its full potential.

%
%
\vspace{0.5cm} AMM thanks the SOC for the invitation and
acknowledges support from Fapesp (grants 97/11299-2 and
98/04267-0) and CNPq. CVR has received financial support from
Fapesp (grants 98/1443-1 and 92/1812-0).

%

\end{document}